\documentclass{egpubl}
\usepackage{eg2022}
 
\ConferencePaper        
\usepackage[T1]{fontenc}
\usepackage{dfadobe}  
\usepackage{pifont}
\newcommand{\cmark}{\ding{51}}%
\newcommand{\xmark}{\ding{55}}%
\biberVersion
\BibtexOrBiblatex
\usepackage[backend=biber,bibstyle=EG,citestyle=alphabetic,backref=true]{biblatex} 
\electronicVersion
\PrintedOrElectronic

\ifpdf \usepackage[pdftex]{graphicx} \pdfcompresslevel=9
\else \usepackage[dvips]{graphicx} \fi

\usepackage{egweblnk} 

\title[Z2P: Instant Visualization of Point Clouds]%
      {Z2P: Instant Visualization of Point Clouds}

\author[\centering G. Metzer,
        R. Hanocka, R. Giryes, N. J. Mitra, D. Cohen-Or]
{\parbox{\textwidth}{\centering G. Metzer$^{1}$,
        R. Hanocka$^{4}$, R. Giryes$^{1}$, N. J. Mitra$^{2, 3}$, D. Cohen-Or$^{1}$ 
        }
        \\
{\parbox{\textwidth}{\centering $^1$Tel Aviv University\\
         $^2$ University College London \\
        $^3$ Adobe Research \\
        $^4$ University of Chicago \\
       }
}
}

\usepackage{hyperref}
\usepackage{amsmath}
\usepackage{graphicx}
\usepackage{comment}
\usepackage{multirow,bigdelim}
\usepackage{array}
\usepackage{wrapfig}
\usepackage{amsfonts}

\usepackage{csvsimple}
\usepackage{adjustbox}
\usepackage[percent]{overpic}

\newcommand{\name}{Z2P\xspace}

\usepackage{dsfont}

\usepackage{pgfplots}
\pgfplotsset{compat=newest}
\usepgfplotslibrary{groupplots}
\usepgfplotslibrary{dateplot}

\usepackage[bottom]{footmisc}
\usepackage{color}
\usepackage{soul}
\usepackage{breqn}
\usepackage{booktabs}
\usepackage{tabularx}
\usepackage{rotating}
\usepackage{xspace}

\definecolor{purple}{rgb}{0.65,0,0.65}
\definecolor{blue}{rgb}{0, 0.2, 0.8}
\definecolor{orange}{rgb}{0.6, 0.6, 0}
\definecolor{red}{rgb}{0.8, 0.2, 0.2}
\definecolor{magenta}{rgb}{0.5, 0.0, 1.0}
\definecolor{black}{rgb}{0.0, 0.0, 0.0}
\definecolor{cyan}{rgb}{0, 0.65, 0.65}
\definecolor{olive}{rgb}{0.2, 0.6, 0.5}
\usepackage{xcolor}

\DeclareRobustCommand{\citet}[1]{\citeauthor*{#1}~\cite{#1}}

\newif\ifdraft
\draftfalse

\ifdraft
\newcommand{\rh}[1]{{\color{olive}\textbf{Rana:} #1}}

\newcommand{\gmc}[1]{{\color{orange}\textbf{Gal:} #1}}
\newcommand{\rgc}[1]{{\color{magenta}\textbf{Raja:} #1}}

\newcommand{\nm}[1]{{\color{violet}\textbf{Niloy:} #1}}

\newcommand{\dcc}[1]{{\color{purple}\textbf{Danny:} #1}}

\newcommand{\revise}[1]{{\color{black}#1}}
\newcommand{\revisee}[1]{{\color{black}#1}}

\else
\newcommand{\rh}[1]{}

\newcommand{\gmc}[1]{}

\newcommand{\nm}[1]{}

\newcommand{\dcc}[1]{}

\newcommand{\rgc}[1]{{}}

\newcommand{\revise}[1]{{\color{black}#1}}
\newcommand{\revisee}[1]{{\color{black}#1}}
\fi

\newcommand{\specialcell}[2][c]{%
  \begin{tabular}[#1]{@{}c@{}}#2\end{tabular}}

\begin{document}

\teaser{
\centering
    \includegraphics[width=\textwidth]{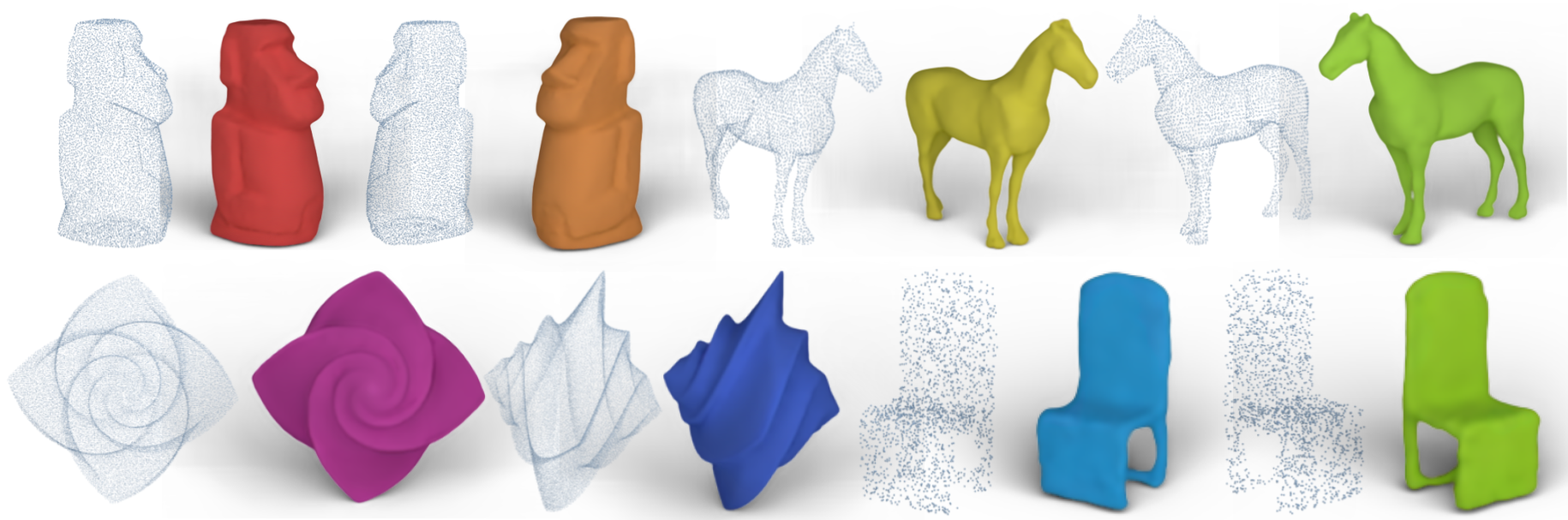} 
    \caption{Point cloud inputs and resulting visualization using our \name, conditioned on different colors and lighting positions.
    }
\label{fig:teaser}
}

\maketitle
\begin{abstract}
   
We present a technique for visualizing point clouds using a neural network. Our technique allows for an instant \textit{preview} of any point cloud, and bypasses the notoriously difficult surface reconstruction problem or the need to estimate oriented normals for splat-based rendering.
We cast the preview problem as a conditional image-to-image translation task, and design a neural network that translates point depth-map directly into an image, where the point cloud is visualized as though a surface was reconstructed from it. Furthermore, the resulting appearance of the visualized point cloud can be, optionally, conditioned on simple control variables (e.g., color and light). We demonstrate that our technique instantly produces plausible images, and can, on-the-fly effectively handle noise, non-uniform sampling, and thin surfaces sheets.




\printccsdesc   
\end{abstract}  
\section{Introduction}

Point clouds are a popular and flexible representation of 3D shapes. A slew of works have successfully employed deep neural networks to synthesize point clouds, for shape synthesis~\cite{achlioptas2018learning, li2018point, yang2019pointflow, hertz2020pointgmm, gal2020mrgan}, upsampling~\cite{yu2018pu, yifan2019patch}, consolidation~\cite{yu2018ec, metzer2020self}, shape completion~\cite{yuan2018pcn}, denoising~\cite{rakotosaona2020pointcleannet}, among others. Thus, generating an increasing demand for a fast and effective technique to \textit{visualize} point clouds. Yet, neurally synthesized point clouds do not contain a globally consistent normal orientation (since standard loss functions do not trivially enable normal regression~\cite{metzer2021orienting}), which is a prerequisite for rendering them \revisee{using surface reconstruction}.

Since points are zero-dimensional entities, they cannot be rendered directly. One approach enables point cloud visualization by converting each point into a local tangent patch (\emph{i.e.,} splats)~\cite{zwicker2001surface, botsch2003high}. Since point samples are irregular and unstructured, different heuristics for deciding the size of local splats inevitably lead to holes and overlaps.
\revisee{Moreover, these techniques often assume normals are provided as input, or ought to be estimated on-the-fly~\cite{diankov2007real, preiner2012auto}. One approach for avoiding the difficult normal estimation problem is through screen space operators~\cite{pintus2011real}, at the expense of limited shading and visualization abilities.}

A more meticulous approach is to first reconstruct a mesh surface from the input point samples~\cite{kazhdan2006poisson, kazhdan2013screened, Hanocka2020point} and then render the mesh. However, reconstructing a surface is a notoriously difficult problem. Such an approach appears to be \textit{overkill} if the end goal is to simply visualize the point set. Further, it inherits the problems of any reconstruction algorithm. For example, neural synthesized point clouds lack normal orientation and often contain interior points, resulting in poor reconstruction results. Moreover, surface reconstruction is unstable and not well defined in the case of thin surfaces and sheets. \revisee{Finally, these methods are often slow, which is not feasible for a quick preview visualization.}
\revisee{One exception is~\cite{katz2007direct}, which allows for faster reconstruction without normal information, but has limited success on sparse point clouds (see Figure~\ref{fig:vase_noise}).}

We cast the problem as an image-to-image translation problem whereby z-buffers, \textit{i.e.}, depth-augmented point features as viewed from target camera view, are directly translated by a neural network to rendered images, conditioned on control variables (e.g., color, light). Once trained, our network produces visualization using a  single forward pass, and hence is instantaneous. 
Further, we support scene controls, such as light position and colors as an input condition to the network using adaptive instance normalization. 
The convolutional nature of the network enables learning a mapping at the patch level, which is location invariant. Yet, the shadows in the scene do depend on the location. Thus, by leveraging a positional encoding, we break this invariance and allow the shading effects to depend on the location.

Our network is trained on automatically generated pairs of point clouds and rendered meshes, and only requires training on a few unique meshes to achieve generalization.
We generate a copious amount of paired data from each mesh, by applying random lighting, color, and rotation augmentations. During inference our method can generate plausible images from point clouds where the ground-truth is unknown (\emph{i.e.,} noisy point clouds from neural networks or scanning devices), bypassing the challenging reconstruction or point normal orientation problem altogether. Moreover, it is significantly faster than the rendering engine it was trained on: our preview visualization takes under one second, while the full-blown rendering engine takes over 10 seconds.

\begin{figure}[h!]
    \centering
    \newcommand{\pl}{-3.7}
    \begin{overpic}[width=\columnwidth]{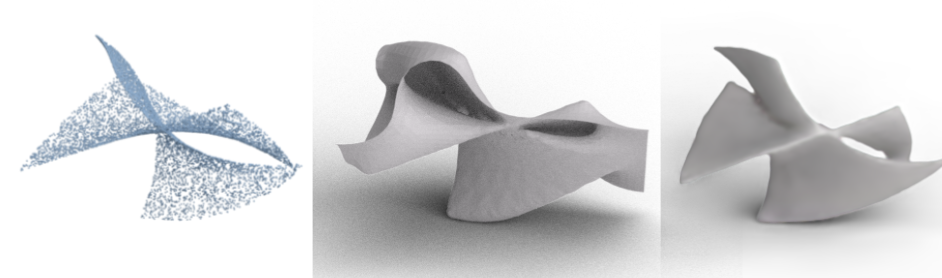}
    \put(8,  \pl){\textcolor{black}{Input}}
    \put(43,  \pl){\textcolor{black}{Poisson}}
    \put(80,  \pl){\textcolor{black}{Ours}}
    \end{overpic}
    \vspace{0.01in}
    \caption{Sheets are especially hard to orient, reconstruct, and may be unprintable. Popular techniques such as Poisson reconstruction~\cite{kazhdan2006poisson} are not well defined in these cases (due to the open surface and poor normals). \rgc{Maybe add here also a comment on neural rendering and splatting and show the result of the latter also in this figure?}
    Our method, which operates on raw point clouds, can handle such cases as it does not rely on oriented normals or other surface attributes.}
    \label{fig:sheet}
\end{figure}

We show that our method can act as an effective, quick-and-dirty point cloud preview, while still providing useful control. We demonstrate that our approach can successfully handle noise, non-uniform sampling, and in particular thin surfaces and sheets (see Figure~\ref{fig:sheet}). Our experimental evaluations show that our technique is comparable or better than existing techniques, and is considerably faster.

\section{Related Work}

\begin{figure}[t!]
    \centering
    \begin{overpic}[width=\columnwidth]{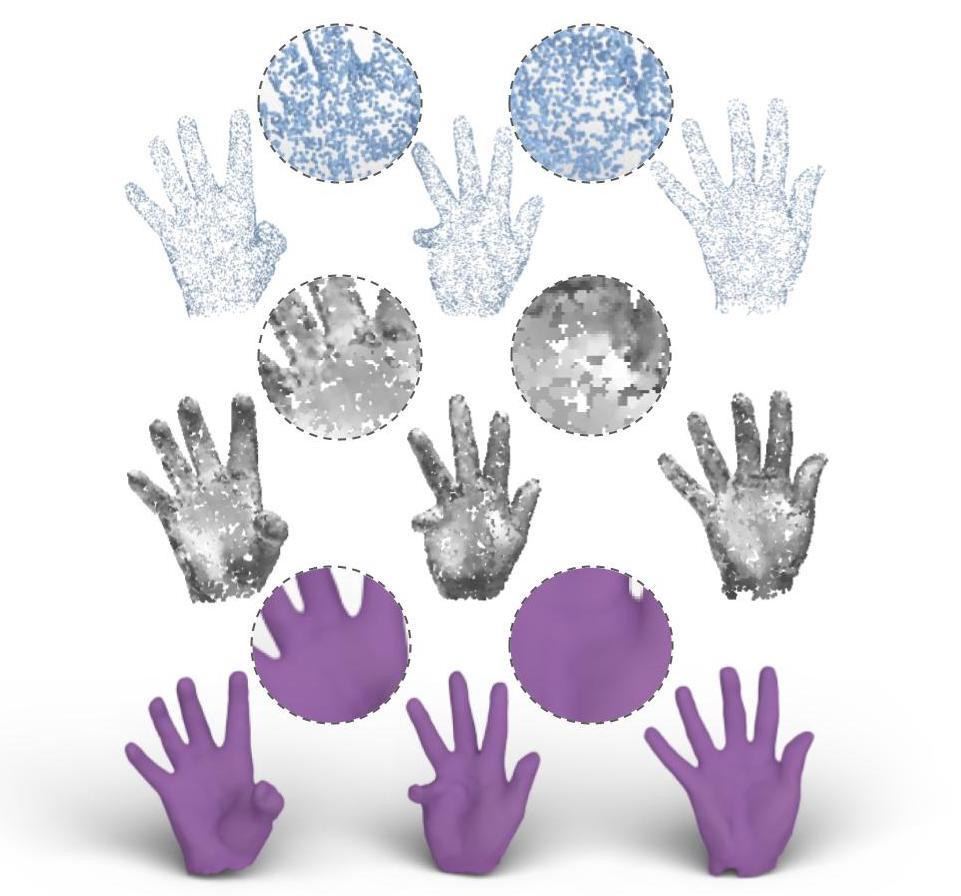}

    \put(0, 6){\rotatebox{90}{Ours}}
    \put(0, 40){\rotatebox{90}{DSS}}
    \put(0, 70){\rotatebox{90}{Input}}
    \end{overpic}
    
    \caption{Our method can directly visualize point clouds with non-uniform sampling. Note the self shadow effects on the palm of the left example.}
    \label{fig:hands}
\end{figure}

\revise{
\textbf{Splatting.} The most common and direct way to render point clouds is through splatting, which are small planes placed at each point in the input point cloud, and point in the normal direction~\cite{zwicker2001surface, diankov2007real, preiner2012auto}. \revisee{Splatting methods generally require normals as input in order to set the splat direction. Some splatting methods such as~\cite{diankov2007real, preiner2012auto} can estimate the normal direction in screen space, instead of relying on explicit normals as input.
Both~\cite{diankov2007real, preiner2012auto} are able to work in real-time and utilize a GPU.}
The orientation of the normal vector can also be considered for shading purposes, to detect back facing points.
Results of different splatting methods can be seen in Figures~\ref{fig:hands}, \ref{fig:vase_noise}, \ref{fig:chairs_shapegf}. Splatting-based methods are adequate for fast and simple visualizations, but may be limited in some features as described in Table~\ref{tab:comparison_to_other_methods}.
}

\revisee{
\textbf{Screen Space Operators.}
\citet{pintus2011real} developed a screen space technique that is able to render point clouds in real-time without normals. 
To render the point clouds, the points are first projected onto a texture, then the method detects the visible points and interpolates the depth values over the rest of the screen. Screen space deferred shading is applied to highlight surface curvature. 
Although this method is real-time and does not require normals, it can only perform deferred shading. 
}

\textbf{Surface reconstruction.} Reconstructing a surface mesh from point clouds is a long standing problem in computer graphics that has been studied extensively for over 20 years~\cite{berger2017survey}. A common approach triangulates a set of points~\cite{edelsbrunner1994three, bernardini1999ball}. Alternatively, an implicit function can be built from the point cloud and normals, and the surface mesh is extracted by finding the zero-crossings~\cite{hoppe1992surface, kazhdan2006poisson, kazhdan2013screened}. Surface reconstruction can also be obtained using a mesh deformation technique, where an initial mesh is incrementally displaced to fit an input point cloud~\cite{sharf2006competing,Hanocka2020point}. 

A common prerequisite for surface reconstruction is calculating a globally consistent normal orientation~\cite{mitra2003estimating, liu2010orienting,GuerreroEtAl:PCPNet:EG:2018, schertler2017towards, jakob2019parallel}. Though some techniques, such as~\cite{mullen2010signing, Hanocka2020point} can reconstruct surfaces with unoriented normals. Triangulation techniques (such as ball-pivot~\cite{bernardini1999ball}), handle thin sheets well, yet inevitably lead to undesirable holes in the reconstruction. On the other hand, implicit surface reconstruction (such as Poisson \cite{kazhdan2006poisson, kazhdan2013screened}) tend to handle holes better, yet occupancy is not-well defined in the case of sheets. Finally, this route is computationally expensive since it requires both reconstructing a surface and then rendering an image.

\revise{\citet{katz2007direct} developed a method to detect visible points from a raw point cloud without normals.
As a by product of the the hidden point removal (HPR), the method also allows for a "quick-and-dirty" single view reconstruction \textit{i.e.} a partial mesh that constrains a surface only from the view direction.
A comparison to \citet{katz2007direct} can be found in Figure~\ref{fig:vase_noise} and in the supplementary material.}
 
\textbf{Deep rendering.} 
Since the introduction of the rendering equation~\cite{kajiya1986rendering}, many works have proposed techniques to speedup the rendering process, for example, using \textit{deferred shading}~\cite{saito1990comprehensible} to produce a render from 2D G-buffers (\emph{e.g.,} depth image, normal maps, etc). 

Although deep rendering works and our work both aim at producing an image from a 3D object from a single view, rendering works have a different focus. Rendering methods usually assume a high quality description of the 3D geometry that can be queried, like continuous surfaces, and focus on producing highly realistic visualizations while supporting controls like lighting, materials, etc. 
Our work, on the other hand, enables a quick preview of the shape (bypassing the surface reconstruction problem), which does not qualify as a full fledged rendering engine.
Recently, deep learning has been used to perform deferred shading~\cite{nalbach2017deepshading} which calculates the G-buffer from the complete 3D scene. However, estimating these maps (\emph{e.g.,} occlusions, light and normal information) is non-trivial for point clouds. RenderNet~\cite{nguyen2018rendernet} learns to produce 2D images from a 3D voxel representation of the shape. Adopting this approach to point clouds is wasteful as it requires creating a voxelized volume (reducing the resolution down to $64 \times 64 \times 64$), as well as require oriented normals.
\citet{hermosilla2019deep} learns to shade surfaces using deep neural networks and G-buffers, which is a faster alternative than path-tracing. Note that we do not have access to G-buffer information in our case.

Neural point cloud rendering has been studied in~\cite{kolos2020transpr, dai2020neural, aliev2019neural}.
These works mainly focus on rendering novel views from a single scene using a sparse set of input views.
In this setting, since a descriptor is optimized for each point, it does not extend to point clouds outside of the training set (single scene). 
This type of problem is different than ours, as we do not aim at visualizing a single scene, instead our technique is generalize to unseen point clouds.
Moreover, our system achieves complete disentanglement and control of lighting, color, and ambient shadows, 
as compared to these works that receive shading and color information from the given RGB images of the scene. A comparison between our work and alternative methods is depicted in Table~\ref{tab:comparison_to_other_methods}.
Recently,~\cite{chen2021ngp} proposed a controllable neural rendering pipeline for neural 3D shapes represented by neural implicits.

An orthogonal line of works propose differential rendering~\cite{yifan2019differentiable, chen2019dibrender} techniques for a plug-and-play module in deep learning architectures. Different than the objective of this work, these works are used to perform shape reconstruction or inverse rendering (\emph{i.e.,} recovery of unknown 3D scene and lighting).

\begin{table*}[]
    \centering
\begin{tabularx}{5.9in}{r||c|c|c|c|c|c} %
\specialcell{} & \specialcell{Ray Tracing} & \specialcell{Splatting} & \revisee{\specialcell{Screen Space}} & \specialcell{Reconstruction} & \specialcell{NPR} & \specialcell{Ours}\\
\hline\hline %
\specialcell{Raw xyz points} & \xmark & \cmark & \cmark & \cmark & \xmark & \cmark \\
\hline
\specialcell{Handles points w/o normals} & - & \revisee{\cmark / \xmark} & \cmark  & \xmark & \cmark & \cmark \\
\hline
\specialcell{Generalizes to unseen objects} & \cmark & \cmark & \cmark & \cmark & \xmark & \cmark \\
\hline
\specialcell{Robust to internal points} & - & \cmark & \cmark & \xmark & \cmark & \cmark \\
\hline
\specialcell{Robust to number of points} & - & \xmark & \cmark & \xmark & \xmark & \cmark \\
\hline
\specialcell{Shadows} & \cmark & \xmark & \xmark & \cmark & - & \cmark \\
\hline
\specialcell{Per point colors} & - & \cmark & \cmark & \cmark & \cmark & \xmark \\
\hline
\specialcell{Scene level scans} & - & \cmark & \cmark & \cmark & \cmark & \xmark \\
[1ex]
\hline
\end{tabularx}
\caption{Comparing Z2P to other visualization techniques.
\textit{Reconstruction} refers to reconstructing a mesh and rendering it with ray tracing,
and neural point rendering \revisee{(NPR)} refers to the prior works of~\cite{kolos2020transpr, dai2020neural, aliev2019neural}. 
\revisee{Splatting contains \cmark / \xmark for requiring normals, as some works such as~\cite{diankov2007real, preiner2012auto} are able to estimate normals on the fly, and therefore technically do not require normals as input. Screen Space refers to the work of~\cite{pintus2011real}, as discussed in the related work it should be noted that this work is limited to deferred shading.}}
    \label{tab:comparison_to_other_methods}
\end{table*}

\section{Method}
We formulate the visualization task as an image-to-image translation from z-buffers of projected point clouds to \textit{surface} images, conditioned on the visualization settings. Our framework is simple, yet effective. We automatically generate simulated training data to train a fully-convolutional network to learn a mapping from point cloud z-buffers to rendered mesh images. We inject visualization control through Adaptive Instance Normalization (AdaIN)~\cite{huang2017arbitrary} layers, enabling a disentangled representation of visualization parameters and shape.
\begin{figure}[h]
    \centering
    \includegraphics[width=\columnwidth]{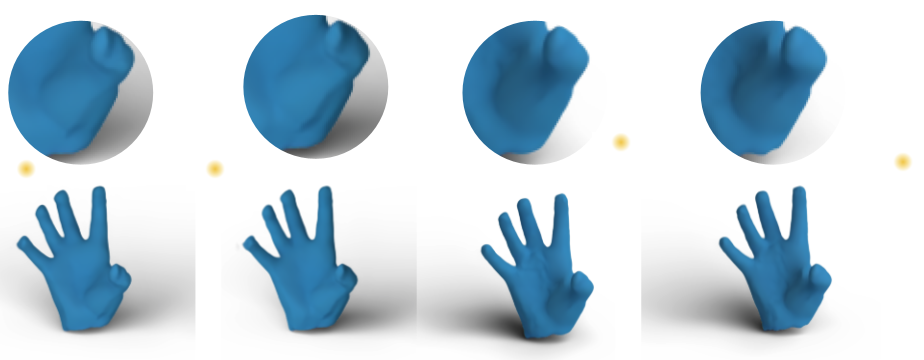}
    \caption{Self shadow effects are plausibly generated with our method. Light position is visualized in yellow. }
    \label{fig:self_shadow}
\end{figure}

Synthesizing shadows requires long range interactions between pixels (see Figures~\ref{fig:self_shadow} and~\ref{fig:ablation_encoding}). This is achieved by the U-Net \cite{ronneberger2015u} structure which takes into account various scales of the input. Yet, in order to enable the network to effectively process the light direction, we break the translational symmetry of the convolutions by appending positional encoding to the input. During inference, we use real point clouds (\emph{i.e.,} noisy point clouds obtained from neural networks or scanning devices) to obtain a plausible quality image, previewed in under a second, bypassing the challenging surface reconstruction or point normal orientation problem altogether.

\begin{figure}[t!]
    \centering
    \includegraphics[width=0.9\columnwidth]{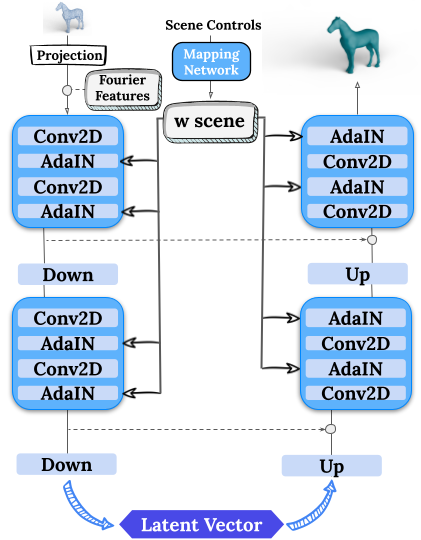}
    \caption{Overview. We project the point cloud into a 2D z buffer image, which is the input to our network. We append positional encoding (Fourier features) in order to model long-range effects required to generate shadows. We inject scene controls into AdaIN after each convolutional layer, enabling control of the light and color in the scene.}
    \label{fig:network_arch}
\end{figure}
\begin{figure*}[h]
    \centering
    \newcommand{\pl}{-1.7}
    \begin{overpic}[width=\textwidth]{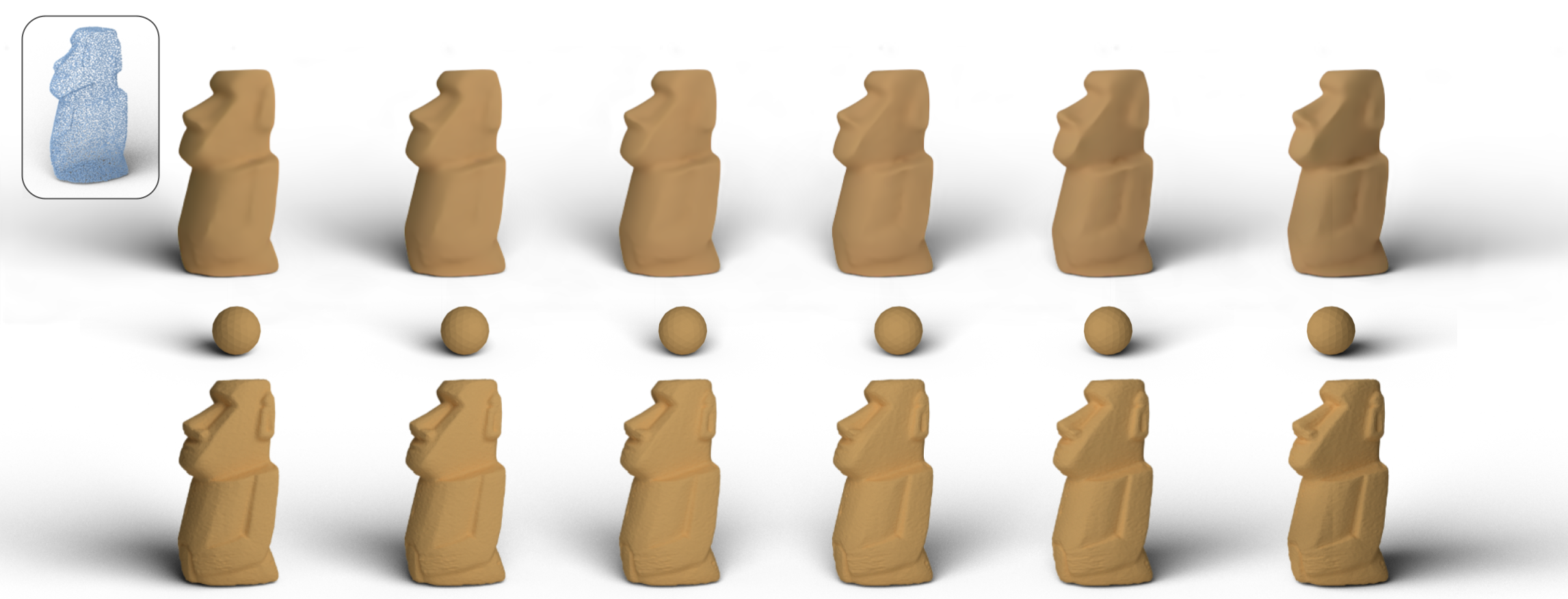}
    \put(\pl, 5){\rotatebox{90}{ground-truth}}
    \put(\pl, 25){\rotatebox{90}{ours}}
    \end{overpic}
    \caption{Lighting control. Top: our visualized result conditioned on slowly changing light position produces a smooth transformation of the shade and objects lighting.
    Bottom: Ground truth mesh renderings under corresponding conditions. The shaded sphere helps visualizing the light direction.}
    \label{fig:shade_control}
\end{figure*}
\subsection{Point Cloud Projection}
We use a 2D z-buffer projection of the point cloud as input to our network. \citet{xie2021style} demonstrated a similar projection was useful for their adversarial loss. Each point in the point cloud $p$ is projected onto the nearest 2D image coordinate $(u_p, v_p)$, with an intensity value proportional to the depth $d_p$, \emph{i.e.,} distance to the image plane. The projection is a perspective transformation according to the camera pinhole model with extrinsic (camera location) and instrinsic (\emph{e.g.,} focal length) parameters. Pixels in the 2D image plane that do not correspond to any 3D points receive an intensity of 0, while other pixels receive an intensity value according to
\begin{equation}
\label{eq:pixel_value}
    z(u_{p}, v_{p})=e^{ -({d_p - \alpha}){/ \beta}},
\end{equation}
where $\alpha$ and $\beta$ are hyperparameters that we fix once throughout training, inference and all experiments. The width of the intensity values for each points' 2D projection on the image is within a $5 \times 5$ window. In the case where multiple points are assigned to the same pixel, we use only the closest one to determine the intensity.

The intuition behind our z-buffer image is that points far away receive a value approaching the background intensity, whereas close points are exponentially brighter than occluded ones. This makes 2D features consistent with the visualization requirements, since far points are usually occluded, which should result in an intensity value that is close to background and much lower than points which are in front of it. Note that our z-buffer image may contain backfacing points, since we do not use oriented point normals. Despite this, we observe that our network is able to cope with such hidden surfaces well.

\subsection{Training Data Generation}
Our training data is automatically generated using Blender 3D engine~\cite{blender}, which provides explicit control over the rendering parameters. We sample meshes to produce corresponding pairs of point cloud and mesh data. In order to generate a large amount of geometric variety using only a small number of unique meshes, we apply random rotations and create different point cloud instances from the same mesh. 
Note that due to the convolutional nature of the network, which shares weights across patches of the input z-buffer, the \emph{effective number} of training examples is larger than the number of pairs of input images.

For each pair of point cloud and underlying mesh pair, we randomly select a color and light location which is used to automatically render an image using Blender Cycles engine. The light is modeled using a flat emission surface with a floor plane for catching shadows and ambient light.

This results in many pairs of training inputs $(z_i, s_i)$ and outputs $r_i$, where $z_i$ is the 2D projection of the point cloud, $s_i$ is a vector containing the settings (color and light position), and $r_i$ is the corresponding blender rendered image. 

The training data is composed of $20$ unique meshes each augmented $400$ times with rotation, color, and lighting position. Each mesh has $10$ corresponding point clouds that were sampled from it. This results in $80K$ pairs of training examples. 

\subsection{Network Architecture}
We formulate our visualization task as an image-to-image translation from z-buffers to rendered images (with an alpha channel for matting). However, this is not a simple image-to-image setup as the target visualization is conditioned on the given color and light position. To enable better control, we propose a modified U-Net~\cite{ronneberger2015u} architecture where we add AdaIN layers to control the color and light position, and use positional encoding to produce accurate shadows. Below, we describe each of these added components. In Section~\ref{sec:exp}, we show the importance of these components through an ablation study.

\begin{figure}[t!]
    \centering
    \includegraphics[width=\columnwidth]{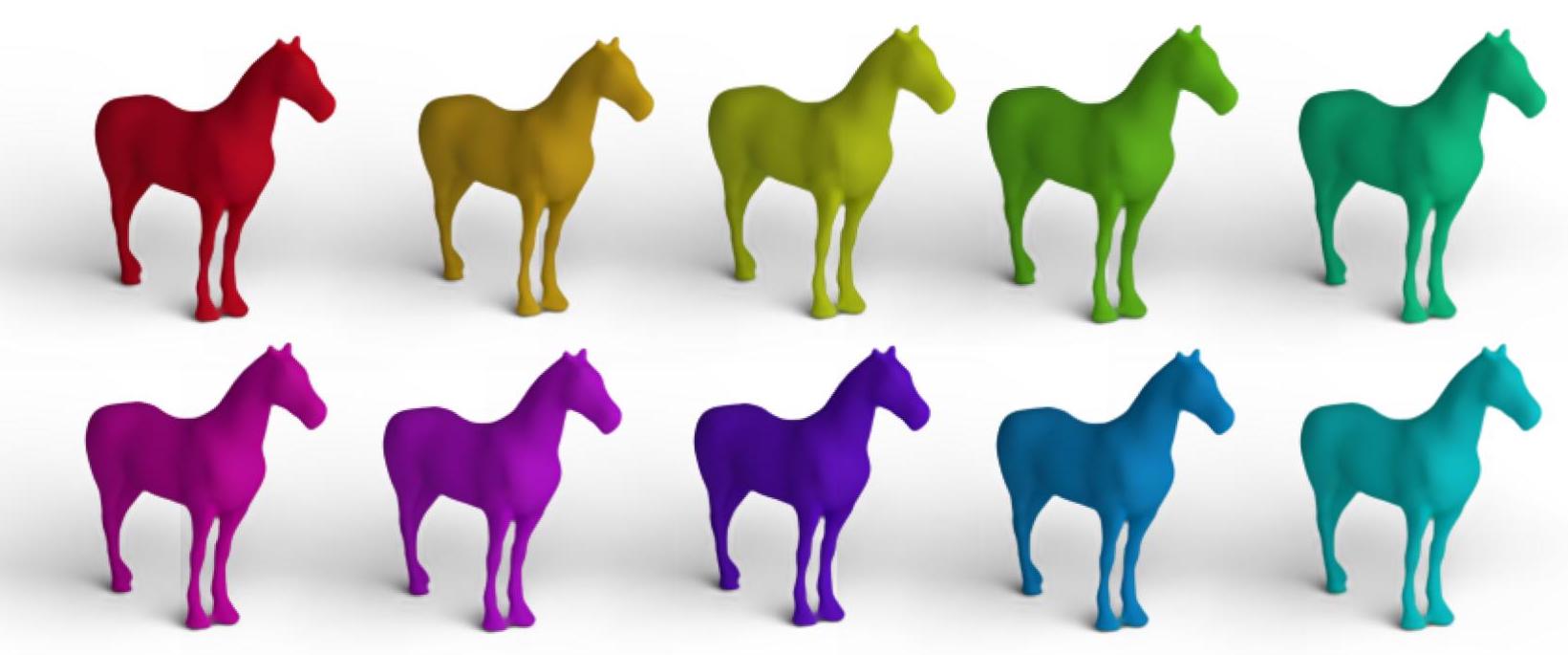}
    \caption{Color control. We keep a constant light and view angle, while continuously varying the input color.}
    \label{fig:control_color}
\end{figure}

\textbf{Visualization control.}
Our architecture enables independent control of both the object color as well as light position of the visualization. This disentanglement can be seen in figures~\ref{fig:control_color} and~\ref{fig:shade_control}.
We inject the color to the network as a 3-dimensional vector, which defines the RGB values of the target shape. 
The light position is given as a 3-dimensional vector corresponding to the light position in spherical coordinates with respect to the camera.
The two vectors are concatenated to produce the settings vector $s\in \mathbb{R}^6$.
We map the settings vector $s\in \mathbb{R}^6$ to a higher dimensional vector $w\in \mathbb{R}^{512}$ by an MLP mapping network.
The vector $w$ is used to produce the  mean and standard deviation that are used by the AdaIN layers. Notice that while $w$ is shared across all AdaIN layers, each of them uses a different (learned) affine transform to control its parameters. This dependence on the light and color through AdaIN allows controlling the scene setting.

\textbf{Adaptive Instance Normalization (AdaIN).}
The original U-Net uses batch normalization after each convolution operation. However, to control the color and shadows of the resulting image, we replace batch normalization with AdaIN layers. These layers normalize the feature map from the prior convolution operation. This was also shown to be an effective technique for controlling global style effects in~\cite{karras2019style}.

Each AdaIN layer $ADA_i$ has a target feature length of size $f_i$. In order to use the global vector $w$, each AdaIN layer contains two additional affine transformations $A_{\beta, i}$, $A_{\gamma, i}$ that receive an input of size $512$ and output a vector of size $f_i$. 

Formally, each AdaIN layer takes as input a feature map $x$ of size $H \times W \times f_i$ and the vector $w$.
The affine transformation is used to calculate
$\beta_i=A_{\beta, i} \cdot w$ and $\gamma_i=A_{\gamma, i} \cdot w$, and the feature map $x$ is normalized according to Equation~\ref{eq:adain} below, where $\mu(x)$ and $\sigma(x)$ are the mean and standard deviation over the spatial dimensions per each channel,
\begin{equation}
\label{eq:adain}
    \hat{x} = \gamma_i \bigg(({x - \mu(x)})/{\sigma(x)}\bigg) + \beta_i.
\end{equation}

\begin{figure}[h]
    \centering
    \newcommand{\pl}{-3.5}
    \newcommand{\pmm}{0}
    \begin{overpic}[width=\columnwidth]{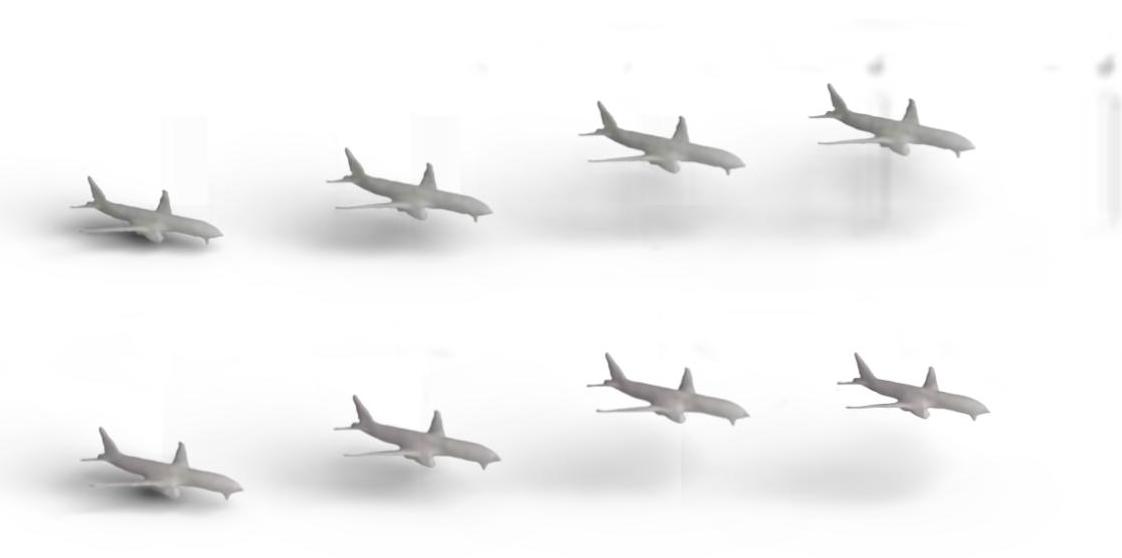}
    \put(\pmm, 5){\rotatebox{90}{w/PosEnc}}
    \put(\pmm, 25){\rotatebox{90}{wo/PosEnc}}
    \end{overpic}
    \caption{Shadow effects require positional encoding. During inference, we move the object location upwards (without having trained on this type of data). Observe how the positional encoding correctly learns the shadow interactions, keeping the floor plane in the correct location.}
    \label{fig:ablation_encoding}
\end{figure}

\textbf{Positional encoding.}
U-Net is a fully convolutional network which means it has a limited receptive field, and is translation equivariant. 
The light position is used to control shadows. This requires long range interactions between pixels, and should not be translation equivariant. For example, when lifting an object in the vertical direction, shadows should remain at the bottom of the image, as the object moves farther away from the bottom floor plane (see Figure~\ref{fig:ablation_encoding}).

To allow for long interactions and break the translational symmetry, we append positional encoding features to the input image that depend on the pixel location. This enables the convolution filters to grasp the global context. We use random Fourier features, which has been successful at modeling both short and long interactions~\cite{tancik2020fourfeat}. 

Formally, each pixel gets a normalized coordinate value $(u_i, v_i)$ in the range  $[0,1]$. At the beginning of training, ten $\{\omega_j\}$ values are sampled from a uniform distribution $\omega_j \sim U[0, 10]$, to produce 40 features for each pixel of the form:
\begin{equation}
enc_i=[\{sin(\omega_j \cdot u_i)\}_j , \{sin(\omega_j \cdot v_i)\}_j, \{cos(\omega_j \cdot u_i) \}_j, \{cos(\omega_j \cdot v_i)\}_j ]
\end{equation}
Each encoding $enc_i$ is concatenated to its corresponding pixel in the z-buffer, which creates additional input features.

\subsection{Losses}
Our network is trained using a combination of simple reconstruction losses. The network produces a 4-dimensional image, three color channels (for RGB) as well as an alpha channel for transparency. We use an $L2$ loss between the network-predicted color image and the ground-truth color image. We also calculate the magnitude of the three color channels (pixel-wise) for both images and calculate the $L1$-norm between them. We also compute the L1 distance between the ground truth and predicted alpha map intensities.

\section{Experiments}
\label{sec:exp}

We performed various experiments to validate our network performance as well as compare to existing surface reconstruction techniques. We demonstrate the ability to generalize on point clouds, where the ground-truth surface is unknown, such as real scans, as well as on point clouds that were synthesized from neural networks. We justify various components of our network architecture and perform run-time comparisons. All results in this paper are shown on held out validation or test set examples. There are extended results and evaluations in the supplementary material and video.

\begin{figure}[b!]
    \centering
    \newcommand{\pl}{-2.7}
    \begin{overpic}[width=\columnwidth]{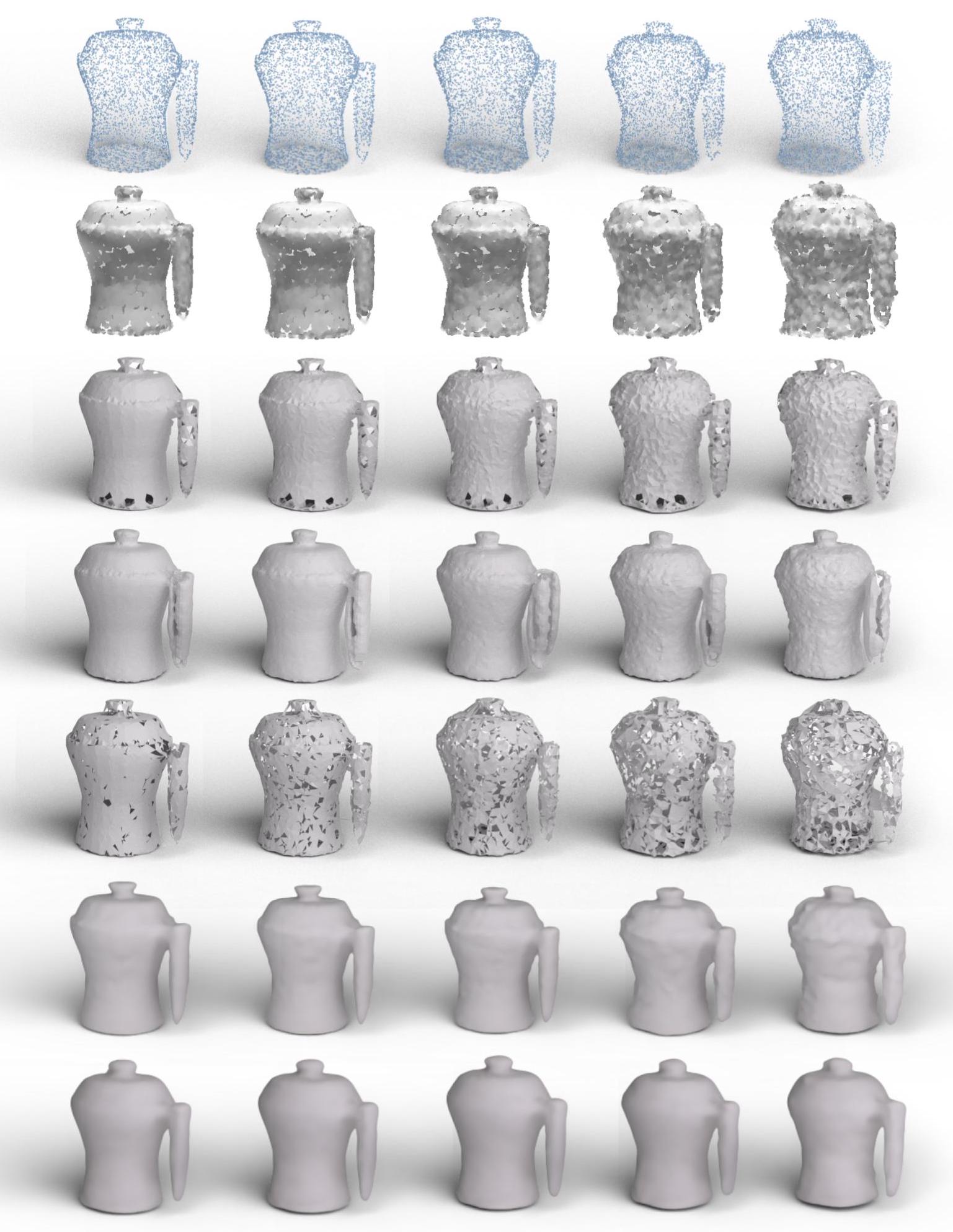}
    \put(8,  \pl){\textcolor{black}{0.001}}
    \put(26,  \pl){\textcolor{black}{0.005}}
    \put(44,  \pl){\textcolor{black}{0.01}}
    \put(60,  \pl){\textcolor{black}{0.02}}
    \put(75,  \pl){\textcolor{black}{0.03}}
    
    \put(0, 0){\rotatebox{90}{Ours Noise}}
    \put(0, 15){\rotatebox{90}{Ours Clean}}
    \put(0, 32){\rotatebox{90}{B-Pivot}}
    \put(0, 46){\rotatebox{90}{Poisson}}
    \put(0, 62){\rotatebox{90}{\revise{HPR}}}
    \put(0, 76){\rotatebox{90}{DSS}}
    \put(0, 89){\rotatebox{90}{Input}}
    \end{overpic}
    \caption{Handling noisy inputs. We compare against different approaches for visualizing a noisy point cloud with increasing levels of Gaussian noise (standard deviation at the bottom). The last row shows results of our network trained with examples corrupted with uniform noise. }
    \label{fig:vase_noise}
\end{figure}

\subsection{Visualization Control}
We control the visualized result by modifying different settings: light position, color, and viewing angle. For example, in Figure~\ref{fig:control_color} (more in supplemental) we control the color of the visualized result. In Figure~\ref{fig:shade_control} we move the light position in the horizontal direction, which changes the ambient shading on the object as well as the shadows on the ground (yet the color is held constant). We also show the visualization result from multiple view points in Figure~\ref{fig:shape_rotation}, while holding light position and color constant. See supplemental video for temporal animations of modifying different visualization controls, as well as the supplemental material.

\subsection{Handling Noisy Inputs}
Depending on the use case, one may want to disregard noise, even if it is part of the input. We trained our model with examples containing uniform noise to produce a type of \textit{denoising} visualizations. To test the ability to disregard noise, we added an increasing level of Gaussian noise during inference (different than the noise we trained on), in Figure~\ref{fig:vase_noise}. The last row contains the results on \name trained with uniform noise, and the second to last row is \name trained on clean data. The noise-removal version visualizes results that are noticeably smoother than the baseline \name, however this may come at the cost of over-smoothing fine details, since they can be attenuated in the presence of noise.

\begin{figure*}[h]
    \centering
    \includegraphics[width=\textwidth]{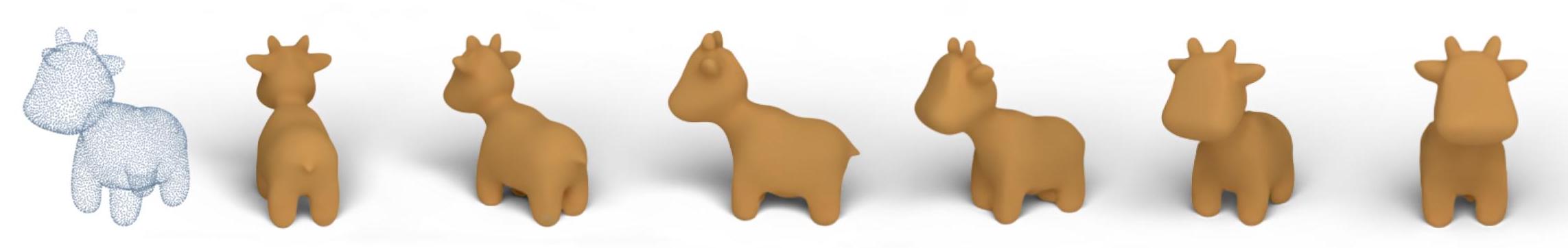}
    \includegraphics[width=\textwidth]{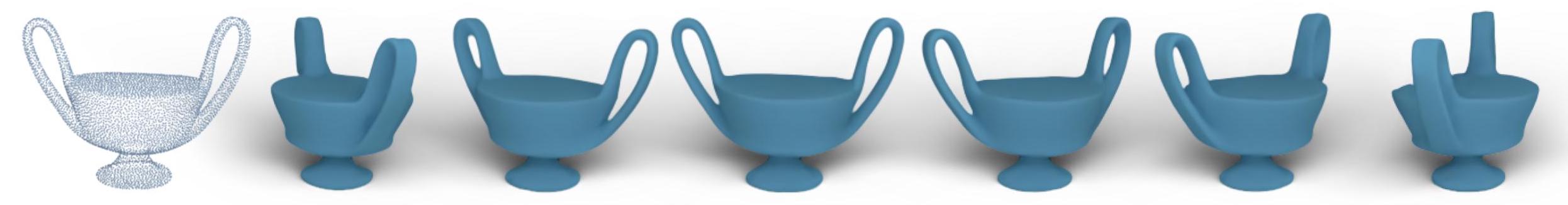}
    \caption{Point clouds can be rendered from multiple views while producing a coherent result with respect to both color and shadows.}
    \label{fig:shape_rotation}
\end{figure*}
\begin{figure}[b!]
    \centering
    \newcommand{\pl}{-3.7}
    \begin{overpic}[width=\columnwidth]{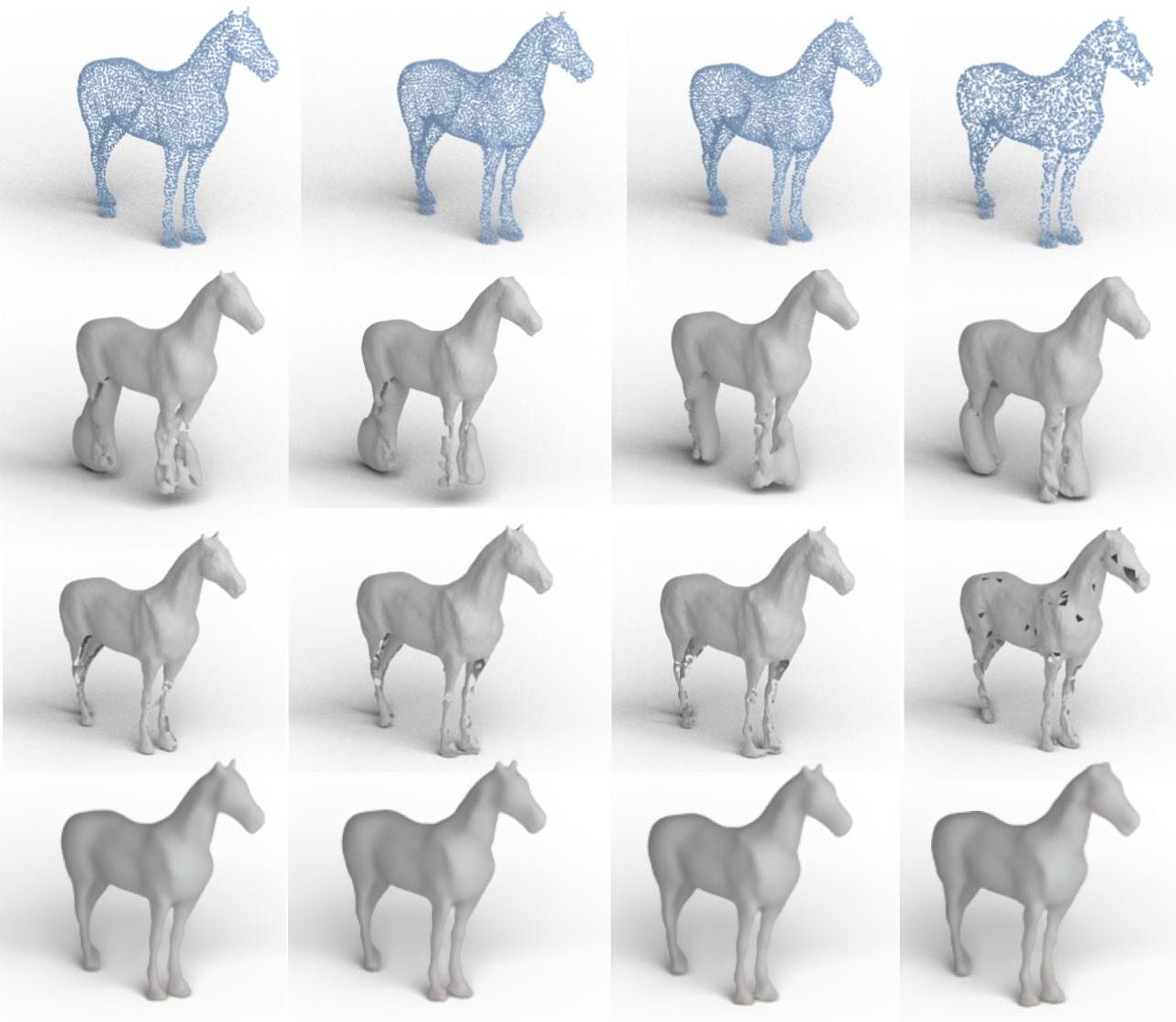}
    \put(8,  \pl){\textcolor{black}{Grid}}
    \put(33,  \pl){\textcolor{black}{FPS}}
    \put(55,  \pl){\textcolor{black}{Poisson Disk}}
    \put(80,  \pl){\textcolor{black}{Uniform}}
    
    \put(0, 6){\rotatebox{90}{Ours}}
    \put(0, 25){\rotatebox{90}{B-Pivot}}
    \put(0, 48){\rotatebox{90}{Poisson}}
    \put(0, 70){\rotatebox{90}{Input}}
    \end{overpic}
\vspace{0.01in}
    \caption{Robustness to different point sampling methods. Poisson reconstruction and ball-pivot (top, middle rows respectively) vary in quality across different sampling methods, our method is able to produce a high quality consistent visualization of the point cloud.}
    \label{fig:horse_sampling_type}
\end{figure}

\subsection{Robustness To Sampling}
We show robustness to different types of sampling methods in Figure~\ref{fig:horse_sampling_type}. Note, that despite only being trained on uniformly sampled meshes, our method is able to generalize to different sampling methods, as well as noise from neural synthesized point networks and real scanning devices.

We also evaluate the ability of our method to cope with point clouds at different scales, which results in a different point density on the z-buffer image in the supplemental material.

\subsection{Generalization}
We demonstrate the ability of our technique to visualize images from point clouds where the ground-truth underlying surface is unknown. Such an example is point clouds synthesized from neural networks. These are interesting since they are challenging for surface reconstruction methods~\cite{metzer2021orienting}, since they do not contain normal information and often have inner points and single sheets. 
In Figure~\ref{fig:chairs_shapegf} and in the supplementary, we show our ability to robustly handle these challenging cases, without oriented normals and with the unknown noise distribution. Future deep learning approaches may use our work to better visualize the generated shapes.

\begin{figure}[h]
    \centering
    \newcommand{\pl}{-1.7}
    \begin{overpic}[width=\columnwidth]{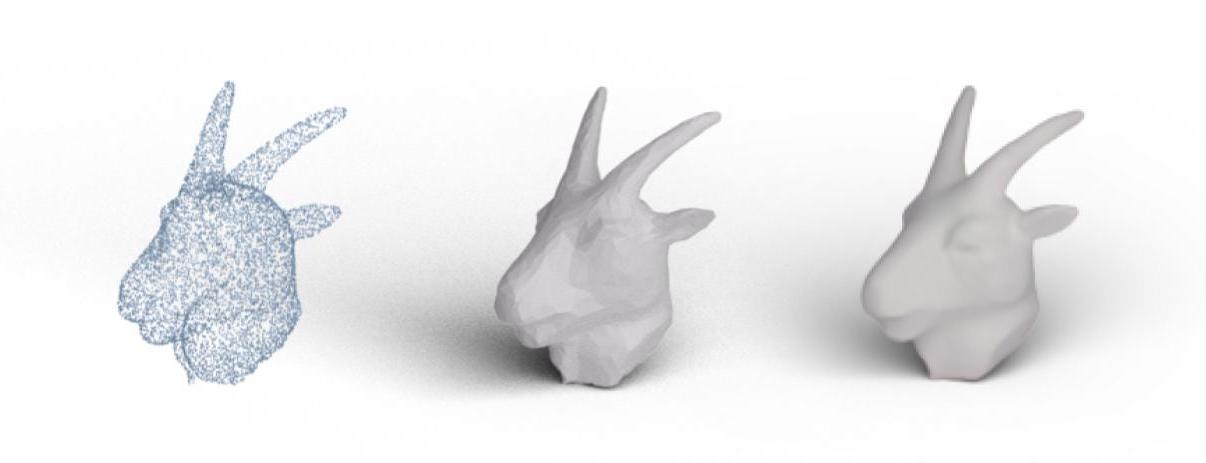}
    \put(15,  \pl){\textcolor{black}{Input}}
    \put(40,  \pl){\textcolor{black}{Point2Mesh}}
    \put(75,  \pl){\textcolor{black}{Ours}}
    \end{overpic}
    \caption{Comparison to \cite{Hanocka2020point}. Though Point2Mesh also produces good results, it requires a long time to converge (30-60 minutes), compared to our result that is obtained in seconds. }
    \label{fig:point2mesh_goat}
\end{figure}
We also compared against Point2Mesh~\cite{Hanocka2020point} which also does not require normal orientation, in Figure~\ref{fig:point2mesh_goat}. Additional comparisons can be found in the supplementary.

\begin{figure}[h]
    \centering
    \newcommand{\pl}{-3.7}
    \begin{overpic}[width=\columnwidth]{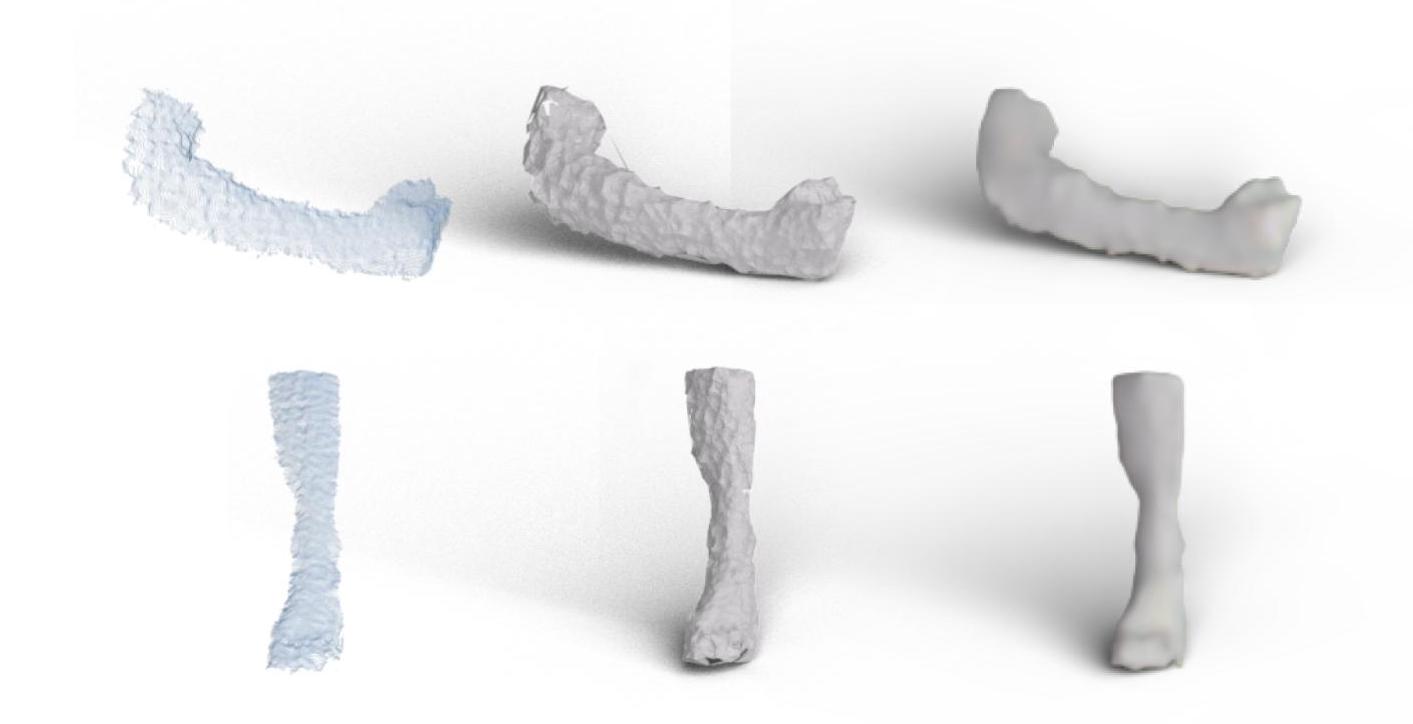}
    \put(20,  \pl){\textcolor{black}{Input}}
    \put(45,  \pl){\textcolor{black}{Ball Pivot}}
    \put(78,  \pl){\textcolor{black}{Ours}}
    \end{overpic}
    \vspace{0.01in}
    \caption{Results of visualizing single view point clouds obtained from Real Sense depth scanner. \rh{TODO swap}}
    \label{fig:single_view_scans}
\end{figure}
\begin{figure}
    \centering
    \newcommand{\pl}{-1.7}
    \begin{overpic}[width=\columnwidth]{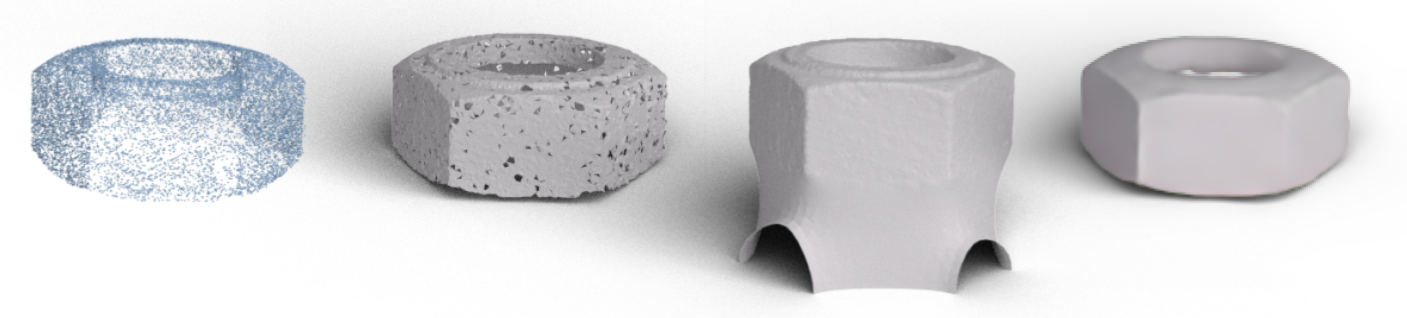}
    \put(8,  \pl){\textcolor{black}{Input}}
        \put(31,  \pl){\textcolor{black}{Ball-Pivoting}}
    \put(58,  \pl){\textcolor{black}{Poisson}}
    \put(83,  \pl){\textcolor{black}{Ours}}
    \end{overpic}
    \caption{Results on real lidar scanned data, which contains holes, noise and open surfaces.}
    \label{fig:screw_scan}
\end{figure}

Scans acquired from sensor devices, are usually obtained as a single occluded view. These scans are not well defined for Poisson reconstruction, but can be used with an explicit triangulation method like ball pivot (which tends to introduce noise). Figure~\ref{fig:single_view_scans} shows results of a single view scan obtained from a low-end \textit{Intel RealSense SR300} scanner. We also show results on a lidar scanner, which contains an open surface which is especially challenging for surface reconstruction techniques, such as Poisson~\ref{fig:screw_scan}.
\begin{figure}[b!]
    \centering
    \newcommand{\pmm}{-0.5}
    \begin{overpic}[width=\columnwidth]{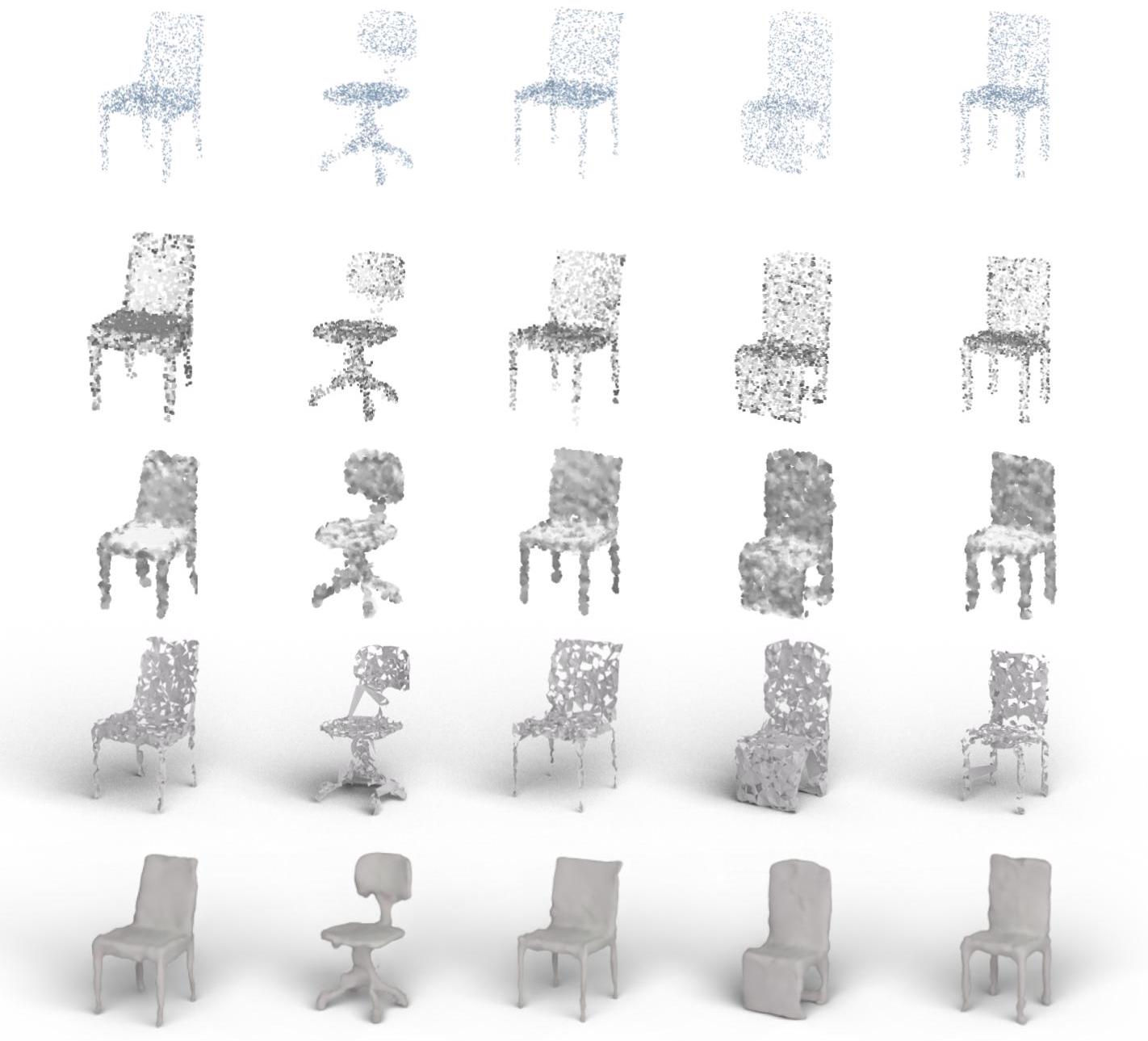}
    \put(\pmm, 5){\rotatebox{90}{Ours}}
    \put(\pmm, 20){\rotatebox{90}{B-Pivot}}
    \put(\pmm, 43){\rotatebox{90}{DSS}}
    \put(\pmm, 56){\rotatebox{90}{MeshLab}}
    \put(\pmm, 80){\rotatebox{90}{Input}}
    \end{overpic}
    \vspace{-1pt}
    \rule[5pt]{0.8\columnwidth}{0.4pt}
    \newcommand{\pl}{-3.7}
    \begin{overpic}[width=\columnwidth]{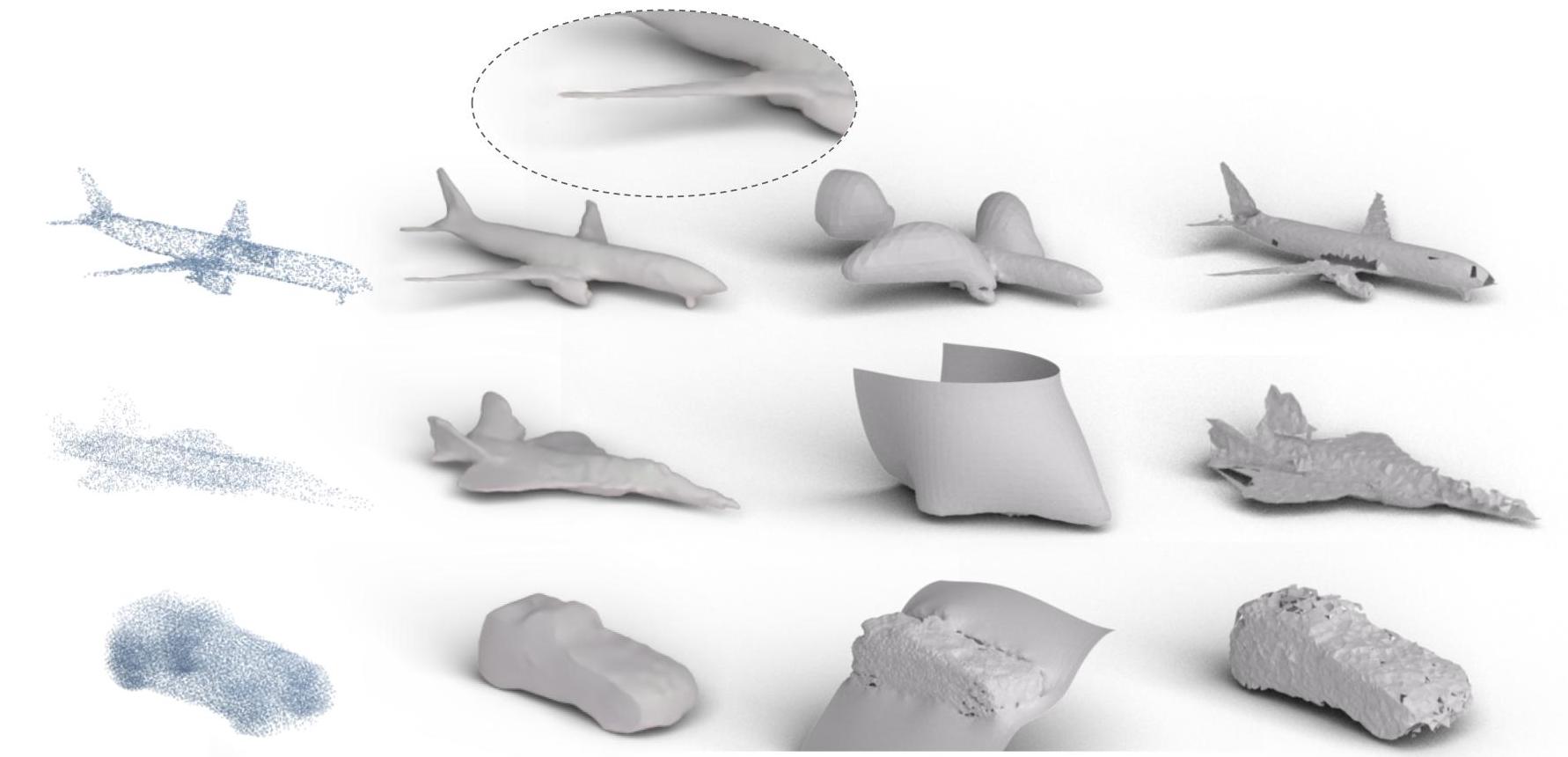}
    \put(8,  \pl){\textcolor{black}{Input}}
    \put(33,  \pl){\textcolor{black}{Ours}}
    \put(57,  \pl){\textcolor{black}{Poisson}}
    \put(80,  \pl){\textcolor{black}{Ball Pivoting}}
    \end{overpic}
    \caption{Our results on different chairs, planes, and cars generated by~\cite{ShapeGF}. Poisson reconstruction was not able to produce a meaningful result for the chairs with thin surfaces in the top row.}
    \label{fig:chairs_shapegf}
\end{figure}

\subsection{Timing Comparison}
Table~\ref{table:runtime_our} shows the run-time of each individual component in our method. The table shows the timings for: projecting the point cloud onto the 2D coordinate plane (\textit{project}), calculating the z-buffer intensity (\textit{z-buffer}), forward pass through the network (\emph{forward}), and the average overall time it takes for one example (\textit{avg per example}). Visualizing a point cloud from start to finish takes under three seconds for a point cloud of size 5k, where the most expensive computation is the z-buffering process. Since the z-buffering code is written in Python, this part can %
be significantly accelerated using GPU and C++. Training our method takes around 24 hours on a single GTX 1080 Ti GPU. Rendering the train and validation set takes 48 hours (using a single GPU with Blender).   

\begin{table}[h]
\centering
 \begin{tabular}{c|c|c|c||c} %
\specialcell{batch \\ size} & \specialcell{project\\ ms} & \specialcell{z-buffer\\ sec} & \specialcell{forward\\ ms} & \specialcell{avg per example \\ sec} \\
\hline\hline %
1 & <0 & 2.60 & 20 & 2.62 \\
2 & 10 & 5.04 & 40 & 2.54 \\
3 & 10 & 7.88 & 60 & 2.65 \\
4 & 10 & 10.70 & 80 & 2.69 \\
5 & 20 & 13.28 & 110 & 2.68 \\
6 & 20 & 17.48 & 120 & 2.93 \\
[1ex] %
\hline %
\end{tabular}
\caption{Run time break down. Visualizing a point cloud from start to finish using our method takes under 3 seconds. The z-buffer operation takes almost all the time, which is written in unoptimized Python code.}
\label{table:runtime_our}
\end{table}

Rendering an image from point clouds by first estimating a reconstruction is computationally expensive. As shown in Table~\ref{table:runtime_surface_blender}, Reconstruction requires estimating the point normals and propagating their orientation ($\sim$3 seconds), running the reconstruction algorithm itself ($\sim$1 or $\sim$10 seconds), and then rendering the result with blender ($\sim$10 seconds). In total, Poisson reconstruction and ball pivot take 13.85 and 23.92 seconds, respectively, whereas our technique on the same point cloud takes less than 3 seconds.

\begin{table}[h]
\centering
\begin{tabular}{l|c|c} %
\specialcell{steps} [sec] & Poisson & B-pivot\\
\hline\hline %
\specialcell{normal estimation} & 2.76 & 2.76 \\
\hline
\specialcell{reconstruction} & 10.63 & 1.03\\
\hline
\specialcell{render} & 10.53 & 10.06 \\
\hline
\specialcell{\textbf{total time}} & \textbf{23.92} & \textbf{13.85}  \\
\hline
\end{tabular}
\caption{Run time break down for surface reconstruction approaches. In total, Poisson reconstruction and ball pivot take 13.85 and 23.92 seconds, respectively, whereas our technique on the same point cloud takes less than three seconds (Table~\ref{table:runtime_our}).}
\label{table:runtime_surface_blender}
\end{table}

\subsection{Architecture Ablation}
\begin{figure}
    \centering
    \newcommand{\pl}{-3.7}
    \newcommand{\pmm}{-3.5}
    \begin{overpic}[width=\columnwidth]{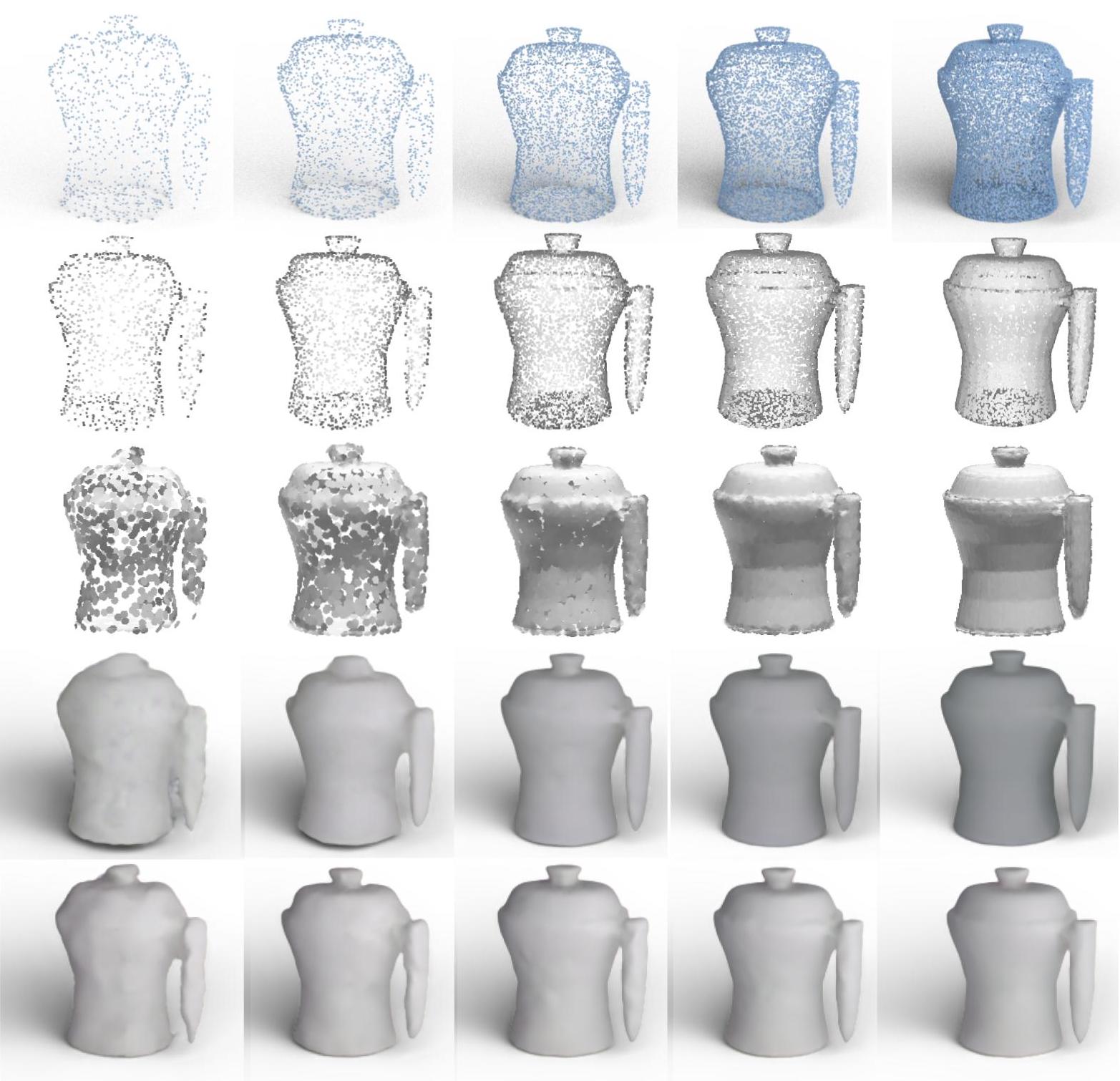}
    \put(8,  \pl){\textcolor{black}{1,000}}
    \put(29,  \pl){\textcolor{black}{2,000}}
    \put(47,  \pl){\textcolor{black}{5,000}}
    \put(64,  \pl){\textcolor{black}{10,000}}
    \put(87,  \pl){\textcolor{black}{20,000}}
    
    \put(\pmm, 0){\rotatebox{90}{AdaIN (Ours)}}
    \put(\pmm, 23){\rotatebox{90}{Features}}
    \put(\pmm, 45){\rotatebox{90}{DSS}}
    \put(\pmm, 59){\rotatebox{90}{MeshLab}}
    \put(\pmm, 80){\rotatebox{90}{Input}}
    \end{overpic}
    \caption{The visualization control vector as extra input features leads to an inconsistent lighting color result with respect to the input density.}
    \label{fig:ablation_density_example}
\end{figure}
The design choices of our architecture are explained through several experiments. We control the visualization through AdaIN, which has been shown to be a powerful tool for controlling style parameters. We compare it to a feature-baseline, where instead of injecting the visualization settings through AdaIN, we create a variant of our network which appends the visualization settings as extra input features (repeated for every input z-buffer pixel). Note how in Figure~\ref{fig:ablation_density_example}, the AdaIN version maintains a consistent color regardless of sampling density. By injecting controls at a global scale with AdaIN, which only changes second order feature map statistics, we obtain a framework with better style consistency and control. 
We also test the impact of the positional encoding. Figure~\ref{fig:ablation_encoding} shows that it is necessary for correctly modeling the long range interactions needed to generate shadows. 

\subsection{Splatting}
In Figures~ \ref{fig:hands}, \ref{fig:chairs_shapegf}, \ref{fig:ablation_density_example} and the Supplementary Material we compare to splatting techniques.
Specifically, we compare to two splatting methods, calculated using MeshLab~\cite{meshlab} and using DSS~\cite{yifan2019differentiable}. We use the forward renderer from DSS. which can be seen as splatting using \citet{zwicker2001surface}.
All splats are rendered with double-sided shading, which is oblivious to the normal orientation. 

As the point clouds used in the paper are relatively sparse, holes are visible in the splatting visualizations.
The sparsity also affects normal estimation, as it is calculated using local neighborhood information, and causes noisy \textit{speckle} artifacts, as seen in Figure~\ref{fig:chairs_shapegf}. 
Splatting techniques are better designed for dense point sets, as can be seen in Figure~\ref{fig:ablation_density_example}, as well as when the points already contain shaded color information.

\begin{figure}[b!]
    \centering
    \newcommand{\pl}{-2.7}
    \begin{overpic}[width=\columnwidth]{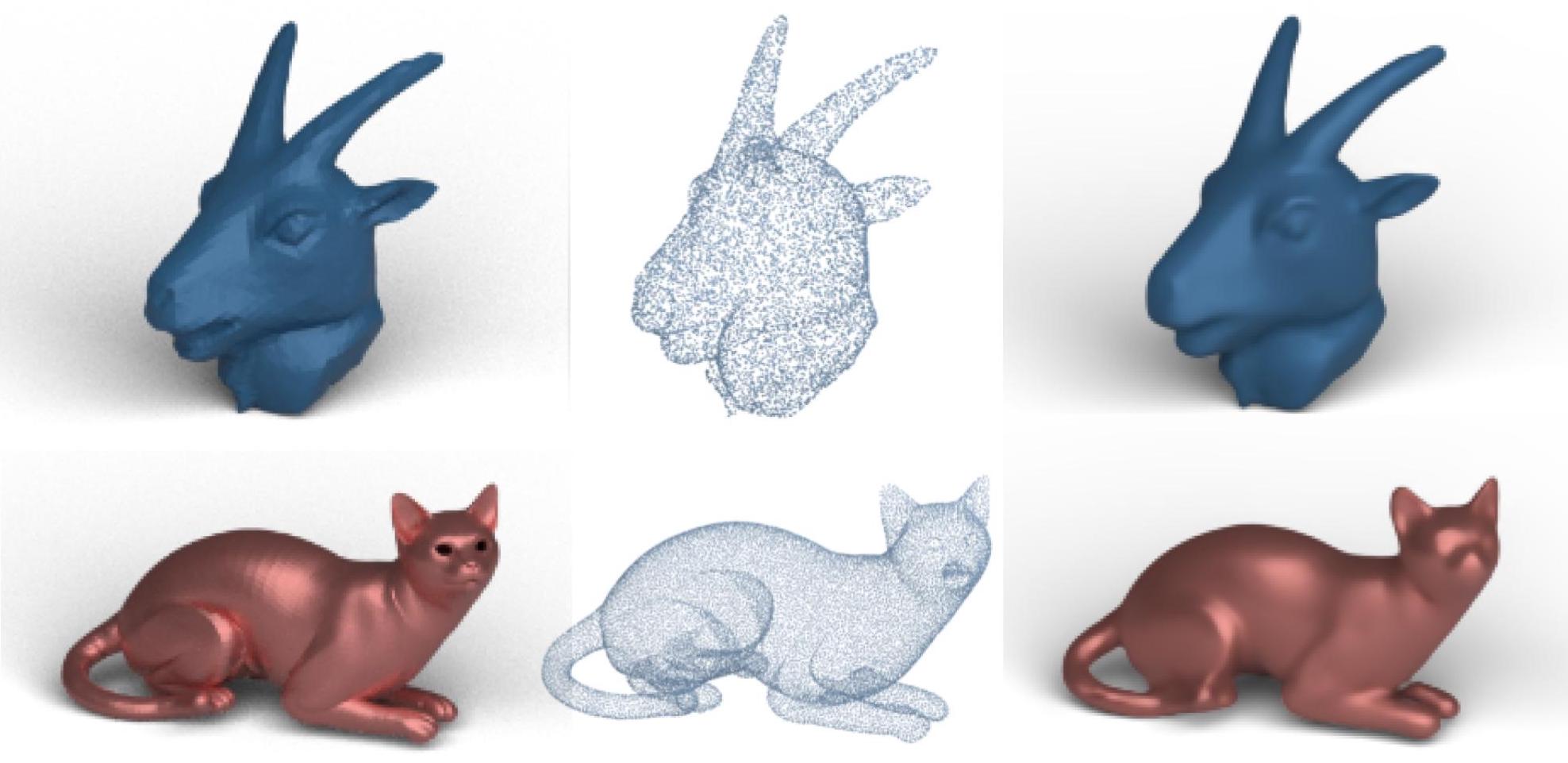}
    \put(7, \pl){\rotatebox{0}{ground-truth}}
    \put(40, \pl){\rotatebox{0}{sampled-pc}}
    \put(79, \pl){\rotatebox{0}{ours}}
    \end{overpic}
    \caption{Two examples of the results of our method, trained on a dataset with the \textit{Principled BSDF} shader. The examples shows that our method can control \textit{Roughness} and \textit{Metallic} attributes. More examples can be seen in the supplementary material.}
    \label{fig:specularity}
\end{figure}
\subsection{Material Control}
All the figures presented throughout the paper, are the results of a model trained with 
the \textit{default} Blender diffuse material, which is sufficient for plausible visualization of point clouds.
To allow for more control over the appearance, we trained our method on a special dataset that uses the Blender \textit{Principled BSDF} shader. In our experiment, we trained the network to control over the dominant \textit{Metallic} and \textit{Roughness} attributes.

In this special dataset, for each training example we sampled random color and light direction controls, as in the original dataset, together with random  \textit{Metallic} and \textit{Roughness} values.
Similar to the basic training method, we input the original controls together with the two additional (\textit{Metallic} and \textit{Roughness}) controls to the network through the AdaIN layers.

Figure~\ref{fig:specularity} shows two examples with different \textit{Metallic} and \textit{Roughness} values as input. The supplementary material contains a sweep over many \textit{Metallic} and \textit{Roughness} values and the corresponding outputs for a better evaluation.

\section{Conclusions and Future Work}

We have presented a neural visualization technique for point clouds. We refer to this technique as \textit{instant visualization} as we bypass the cumbersome step of explicitly converting the point cloud to a continuous surface prior to rendering. Moreover, our instant visualization relaxes the need to compute or use normals; instead, we directly project points to a simple z-buffer. In its basic form, our technique is a \textit{conditional} image-to-image convolutional network.
However, a vanilla fully convolutional network would learn translation at the patch level, and ignore long-distance relations. We showed that by injecting positional encoding, we break the locality and gain non-local shading effects.

\begin{figure}[h!]
    \centering
    \includegraphics[width=\columnwidth]{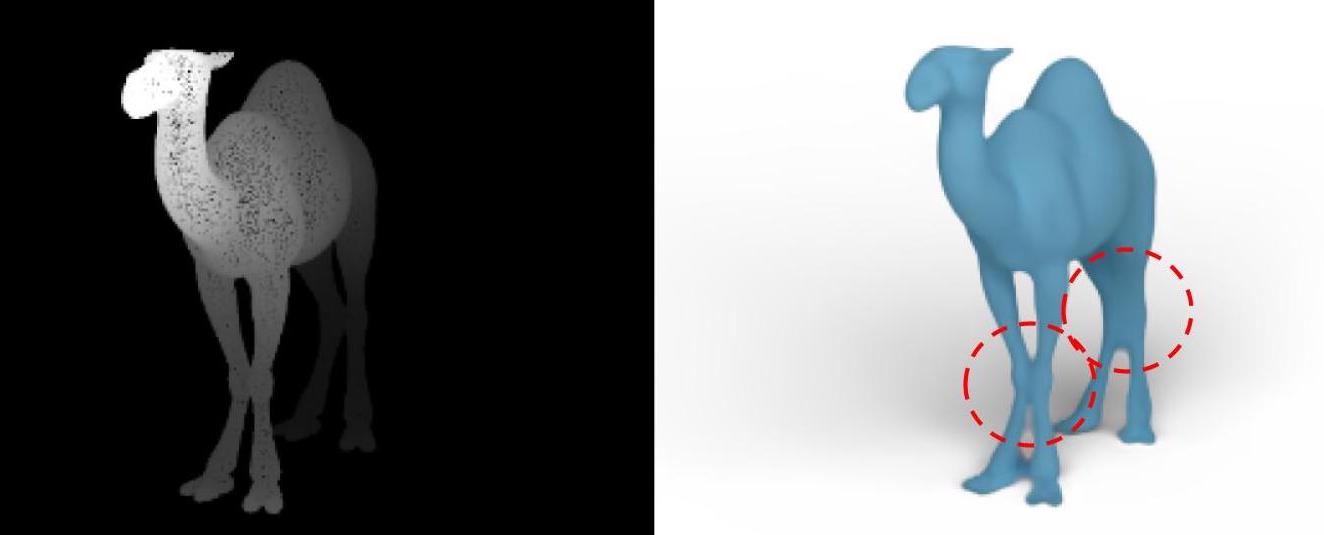}
    \caption{Limitation. Our network may have difficulty distinguishing between nearby surfaces, and incorrectly visualization them as a continuous surface. This problem arises from the ambiguity between typical sampling space and surface holes. }
    \label{fig:limitation}
\end{figure}

\textbf{Limitations and future directions.} Our visualizations may not always be completely accurate, for example, nearby points in Euclidean space may be visualized as touching, when in fact they are far in geodesic proximity (see Figure~\ref{fig:limitation}). \revise{Another problem may arise at fine details, as convolutional neural networks tend to output smooth results.
The smoothing effect is also encouraged by the MSE loss our network is trained with, as this loss aims at achieving the \textit{average} visualization result. 
This problem shows up in Figure~\ref{fig:specularity}, where small holes in the cat's eyes are visualized as closed off.
A dedicated experiment for the robustness to small details, using spherical harmonics with increasing frequency, is shown in the in the supplementary material }

Our framework offers limited control compared to a full rendering engine. While this can partially be addressed by incorporation textures, materials, and other types of lighting models in the learning procedure, our method should not be seen as a replacement for a rendering system in scenarios where physically-based rendering is more important compared to instant visualization.

There is an interesting trade-off between speed and shading locality that we plan to explore more in the future. Local shading effects, i.e., ambient, diffuse and specular can be learned at the patch level, while global shading, like shadows or reflections, are significantly harder and require learning non-local interactions. One research avenue is to incorporate transformers for learning inter-patches dependencies. Another challenging direction is to improve the visualization of fine details including sharp features. These, however, are still challenging also for slower methods that reconstruct surfaces from sparse point clouds, let alone for an instant preview.


\printbibliography                


\end{document}